# IMPLEMENTATION OF THE TRIGONOMETRIC LMS ALGORITHM USING ORIGINAL CORDIC ROTATION

Nasrin Akhter[1], Kaniz Fatema[2], Lilatul Fersouse[3], Faria Khandaker[4]

[1]Department of Computer Science, Stamford University, Dhaka, Bangladesh
nasrin.ah@gmail.com
[2]Department of Computer Science, Stamford University, Dhaka, Bangladesh
shetu789@yahoo.com
[3]Department of Computer Science, Stamford University, Dhaka, Bangladesh
juthi_li@yahoo.com
[4]Department of Computer Science, Stamford University, Dhaka, Bangladesh
faria_cse@yahoo.com

***ABSTRACT***

*The LMS algorithm is one of the most successful adaptive filtering algorithms. It uses the instantaneous value of the square of the error signal as an estimate of the mean-square error (MSE). The LMS algorithm changes (adapts) the filter tap weights so that the error signal is minimized in the mean square sense. In Trigonometric LMS (TLMS) and Hyperbolic LMS (HLMS), two new versions of LMS algorithms, same formulations are performed as in the LMS algorithm with the exception that filter tap weights are now expressed using trigonometric and hyperbolic formulations, in cases for TLMS and HLMS respectively. Hence appears the CORDIC algorithm as it can efficiently perform trigonometric, hyperbolic, linear and logarithmic functions. While hardware-efficient algorithms often exist, the dominance of the software systems has kept those algorithms out of the spotlight. Among these hardware-efficient algorithms, CORDIC is an iterative solution for trigonometric and other transcendental functions. Former researches worked on CORDIC algorithm to observe the convergence behavior of Trigonometric LMS (TLMS) algorithm and obtained a satisfactory result in the context of convergence performance of TLMS algorithm. But previous researches directly used the CORDIC block output in their simulation ignoring the internal step-by-step rotations of the CORDIC processor. This gives rise to a need for verification of the convergence performance of the TLMS algorithm to investigate if it actually performs satisfactorily if implemented with step-by-step CORDIC rotation. This research work has done this job. It focuses on the internal operations of the CORDIC hardware, implements the Trigonometric LMS (TLMS) and Hyperbolic LMS (HLMS) algorithms using actual CORDIC rotations. The obtained simulation results are highly satisfactory and also it shows that convergence behavior of HLMS is much better than TLMS.*

## 1. INTRODUCTION

The current trend back toward hardware intensive signal processing has uncovered a relative lack of understanding of hardware signal processing architectures. Many hardware efficient algorithm exists, but these are not generally well known due to the dominance of software systems over the past quarter century. Among these algorithms is a set of shift-add algorithms collectively known as CORDIC for computing a wide range of functions including certain trigonometric, hyperbolic, linear and logarithms functions. Since the last few years, the implementation and application of CORDIC (COordinate Rotation Digital Computer) arithmetic continue to evolve into very useful areas because of its numerical stability, efficiency in evaluating trigonometric functions and hyperbolic transformations, hardware compactness and computational simplicity [4]. The CORDIC algorithm was originally developed as a digital solution for real time navigation problems. The original work was credited to Jack Volder [3]. Extensions to the CORDIC theory based on work by John Walther [5] and others provide solutions to a broader class of functions. The CORDIC algorithm has found its way into diverse applications including the 8087 math coprocessor [6], the HP-35 calculator, radar signal processors and robotics [7]. In the field of signal processing, CORDIC method has been employed successfully for a wide range of applications like performing FFT, DCT, DST, SVD and other matrix operations, filtering and array processing [3]. For the case of the LMS based adaptive filters, however, the CORDIC based approach has so far remained confined largely to lattice filters [8] because the computations in each stage of the lattice filter can be related easily to a set of hyperbolic operations, while no such direct hyperbolic, or, trigonometric interpretation exists for the computations present in the transversal filter. But two new versions of the LMS algorithm (Trigonometric LMS (TLMS) [1] and Hyperbolic LMS (HLMS) [2]) were proposed which can be realized by CORDIC. In [1] and in [2], The TLMS and the HLMS algorithms were simulated extensively but all of those simulations were performed by exploiting the software-generated output of CORDIC blocks directly. In this paper, the convergence performance of the TLMS and the HLMS algorithms were verified by using internal step-by-step CORDIC rotations for 16 and 32 steps. The obtained simulation results are highly satisfactory and a few of these are shown here. The main scope of the present work is to verify the convergence performance of the TLMS and HLMS algorithms by using internal step-by-step CORDIC rotations. Studies have been made on the internal configuration of the CORDIC block, on the steps or rotations performed by the actual hardware and on the characteristics of TLMS and HLMS algorithms. The simulations performed by the former researchers to observe the convergence behaviour of the two popular algorithms have been studied which used direct output of CORDIC block. Here, in our present work, actual steps performed by CORDIC hardware are considered which was never attempted before. Now the simulations are performed using internal step-by-step CORDIC rotations to observe the actual characteristics of the TLMS and HLMS algorithms. Comparisons have also been made between convergence performance of TLMS and HLMS algorithms using direct output of CORDIC block and convergence performance using step-by-step CORDIC rotations to investigate if they converge at all, if so how well they actually converge.

## 2. REVIEW OF THE LITERATURE

Filtering has become an unavoidable phenomenon in the field of Digital Signal Processing. Data or signal gets distorted or become corrupted as it transmits through a transmission medium before reaching the receiving end because of some impairments such as inter symbol interference (ISI)[10]. In this situation, if the recipient makes any decision based on the received signal, it is most likely that the results would be erroneous. That's why digital data transmission or telecommunication badly needs filtering/equalizing, so that data are to be recovered at the receiving end accurately and efficiently for high speed data transmission. Equalizer [11],[12] is one such functional unit that tries to nullify the adverse effects of ISI on the transmitted signal generally introduced by time dispersive, band limited communication channels. The basic structure of an automatic adaptive equalizer is very much similar to that of an adaptive filter where the weights are adjusted to minimize the average error power at the output with respect to the transmitted symbols. High processing speed in the context of convergence speed can be achieved through two popular approaches: pipelining [13] and parallel processing [14]. In this dissertation, an attempt has been made to develop several new pipelined architectures for a class of adaptive equalizers based on the least mean square (LMS) [15] family of algorithms and to examine in depth various issues pertaining to their implementation. LMS is one of the most well known adaptive algorithms, which updates the filter coefficients by using an approximate version of the steepest descent procedure [16]. Although CORDIC [3] [5] realization of filtering algorithms has been worked upon vastly, research works to improve the performance of filters/equalizers are going on. The fundamental operations in almost all the adaptive equalizers are suitable to be carried out by multiply-accumulate (MAC) [17] units due to the some specific requirement of operation both in filtering as well as in weight updatation. Now-a-days the CORDIC algorithm for realizing filtering algorithms is taking important role on stage of research works in the field of DSP. Since the last two decades, the implementation and application of CORDIC arithmetic continue to evolve into very rich and useful areas as it offers a unified iterative formulation to efficiently evaluate many elementary functions such as trigonometric, hyperbolic, exponential, logarithmic etc [5]. For the case of the LMS based adaptive filters, the CORDIC approach has so far been applied only to the gradient adaptive lattice (GAL) [18] filters, primarily because the computations in each stage of the lattice can be related to a set of hyperbolic rotations. However, more extensive work in this area was carried out by Hu et al [8], who proposed a "CORDIC Adaptive Lattice Filter" (CALF) that directly updates the rotation parameters for each stage instead of the reflection coefficients. Two new versions of the LMS, viz. the trigonometric LMS (TLMS) and the hyperbolic LMS (HLMS) algorithms have been proposed. CORDIC realization of the transversal adaptive filter using trigonometric LMS algorithm was proposed by M. Chakraborty, A. S. Dhar and Suraiya Pervin [1]. Another excellent work was done on CORDIC realization of transversal filters by M. Chakraborty, Suraiya Pervin and T. S. Lamba. They used hyperbolic LMS algorithm in their realization [2]. But both of the works used direct CORDIC output ignoring the internal step-by-step CORDIC rotations. In this situation a question still remains of whether the TLMS and HLMS algorithms perform satisfactorily if the actual steps of the CORDIC hardware is used. In present work, an attempt have been made to implement the TLMS and the HLMS algorithms using the internal step-by-step rotations of the CORDIC hardware. Some comparisons have also been made between the previous work results and the present work results to yield the actual characteristics of the two popular algorithms. As far as present work is concerned, some of the references mentioned may seem loosely linked. However, all the works inspired the present workers to proceed with their work and provided enough resources to realize the details of Digital Signal Processing.

## 3. CORDIC ALGORITHM

CORDIC [3] is an iterative arithmetic algorithm which provides a unified iterative formulation to efficiently evaluate many elementary functions such as trigonometric, hyperbolic, exponential, logarithmic etc. Specifically, all the evaluation tasks in CORDIC are formulated as a rotation of a $2\times1$ vector in various coordinate systems. The basic concept of the CORDIC computation is to decompose the desired rotation angle into the weighted sum of a set of predefined elementary rotation angles such that the rotation through each of them can be accomplished with simple shift-and-add operation. Given a rotation angle $\theta$, it is represented in this method as [4]:

$$\theta \approx \sum_{i=0}^{M-1} \delta_i \alpha_i \qquad (1)$$

where $M$ is the total number of elementary rotation angles and the $i$-th angle $\alpha_i$ is given by:

$$\alpha_i = \frac{1}{\sqrt{m}} \tan^{-1}[\sqrt{m}\, 2^{-s(m,i)}] = \begin{cases} 2^{-s(0,i)} & m \to 0 \\ \tan^{-1} 2^{-s(1,i)} & m = 1 \\ \tanh^{-1} 2^{-s(-1,i)} & m = -1 \end{cases} \qquad (2)$$

In the above equation, $m = 1, -1,$ and 0 correspond to the rotation operation in a circular, a hyperbolic and a linear coordinate system respectively. The norm of a vector $\theta$ in these three coordinate systems is defined as $\sqrt{x^2 + my^2}$. The symbols $\delta_i$, $0 \le i \le M-1$, denote a sequence of $\pm 1$s which determines the direction of each elementary rotation. The integers $s(m,i)$, $0 \le i \le M-1$, form a non-decreasing shift sequence.

Following the above definitions, the basic CORDIC algorithm can be described as follows:

Initiation: Given $x_0$, $y_0$, $z_0$ ( $z_0 = \theta$ ),

/* CORDIC iteration equation */

$$\begin{bmatrix} x_{i+1} \\ y_{i+1} \end{bmatrix} = \begin{bmatrix} 1 & -m\delta_i 2^{-s(m,i)} \\ \delta_i 2^{-s(m,i)} & 1 \end{bmatrix} \begin{bmatrix} x_i \\ y_i \end{bmatrix}, \quad i = 0 \text{ to } M-1, \tag{3}$$

/* Angle updatation unit */

$$z_{i+1} = z_i - \delta_i \alpha_i \tag{4}$$

/* Scaling operation (required for $m = \pm 1$ only) */

$$\begin{bmatrix} x_f \\ y_f \end{bmatrix} = \frac{1}{K_m} \begin{bmatrix} x_M \\ y_M \end{bmatrix} \tag{5}$$

The scaling operation is required to make sure that after rotation, the final coordinate $[x_f \ \ y_f]^T$ has the same *m*-norm as the initial coordinate $[x_0 \ \ y_0]^T$. That is,

$$(x_f)^2 + m(y_f)^2 = x_0^2 + my_0^2, \quad m = -1 \ \ or \ \ +1$$

In conventional format, the CORDIC equations for circular rotation [9] can be written as

$$\alpha_i = \tanh^{-1} 2^{-i}, \tag{6a}$$

$$y_{i+1} = y_i + \delta_i 2^{-i} x_i \tag{6b}$$

$$z_{i+1} = z_i - \delta_i \alpha_i, \tag{6c}$$

where,

$$\alpha_i = \tan^{-1} 2^{-i}, \ i = 0, 1, \ldots, M-1. \tag{7}$$

and for hyperbolic transformation,

$$x_{i+1} = x_i + \delta_i 2^{-i} y_i, \tag{8a}$$

$$y_{i+1} = y_i + \delta_i 2^{-i} x_i \tag{8b}$$

$$\alpha_i \quad = \quad \tanh^{-1} 2^{-i}\,,\ i \quad = 1, \quad 2, \ldots, M-1 \tag{9}$$

The convergence relationship for CORDIC can be portrayed as:

$$\left| z_0 - \sum_{i=0}^{M-1} (\delta_i \alpha_i) \right| \quad \le \quad \alpha_{M-1} \tag{10}$$

This inequality is fulfilled by the CORDIC circular rotation but not by the hyperbolic rotation for the reason that

$$\arctan[2^{-(i+1)}] \quad > \quad (\tfrac{1}{2})\arctan(2^{-i})\,,\ i.e.,\ \text{term}\ (i+1) \quad > \quad \tfrac{1}{2}\ \text{of term}\ i, \tag{11}$$

but

$$\arctan h[2^{-(i+1)}] \quad < \quad (\tfrac{1}{2})\arctan h(2^{-i})\,,\ i.e.,\ \text{term}\ (i+1) \quad < \quad \tfrac{1}{2}\ \text{of term}\ i. \tag{12}$$

For this reason, some "double pass" operations are needed to perform hyperbolic rotations.

Counter clockwise rotation (circular) of a vector $[x_0\ \ y_0]^T$ by an angle $\theta$ can be represented as

$$x \quad = \quad x_0 \cos\theta - y_0 \sin\theta \tag{13}$$

$$y \quad = \quad y_0 \cos\theta + x_0 \sin\theta \tag{14}$$

for $\alpha_i$, $i \quad = \quad 0, 1, \ldots, M-1$, the equations (13) and (14) can be recursively updated as

$$x_{i+1} \quad = \quad x_i \cos\delta_i\alpha_i - y_i \sin\delta_i\alpha_i \tag{15a}$$

$$y_{i+1} \quad = \quad y_i \cos\delta_i\alpha_i + x_i \sin\delta_i\alpha_i\,. \tag{15b}$$

(15a) can be written as:

$$x_{i+1} \quad = \quad \cos\delta_i\alpha_i\,(x_i - y_i \tan\delta_i\,\alpha_i)\,, \tag{16}$$

which reduces to

$$x_{i+1} \quad = \quad \cos\alpha_i\,(x_i - \delta_i 2^{-i} y_i) \tag{17}$$

From (15b), we get,

$$y_{i+1} \quad = \quad \cos\alpha_i\,(y_i + \delta_i 2^{-i} x_i) \tag{18}$$

So, the CORDIC scale factor for circular rotation is

$$K_M \ = \ \prod_{i=0}^{M-1} \cos\alpha_i \quad = \quad \prod_{i=0}^{M-1} \cos(\arctan 2^{-i}) \tag{19}$$

Since it depends on '*M*' only, $K_M$ is a machine constant.

The 32 elementary angles for CORDIC circular rotations are shown in Table-1.

In case of hyperbolic rotation, the scaling factor can be derived as,

$$K_M = \prod_{i=1}^{M} \cosh(\operatorname{arctanh} 2^{-i}) \tag{20}$$

(for $M = 24$, $i$ = 1, 2, 3, 4, 5, 5, 6, 7, 7, 8, 8, 9, 10, 11, 12, 13, 13, 14, 15, 15, 16, 17, 17,18, 19, 19, 20, 20, 21, 21, 22, 22, 23, 24, 24).

Table 1. 32 elementary angles

| | $\tan \theta = 2^{-i}$ |
|---|---|
| $45^0$ | 1 |
| $26.650511771^0$ | 0.5 |
| $14.0362434679^0$ | 0.25 |
| $7.1250163489^0$ | 0.125 |
| $3.576334375^0$ | 0.0625 |
| $1.78991060825^0$ | 0.03125 |
| $0.89517371021^0$ | 0.015625 |
| $0.44761417086^0$ | 0.0078125 |
| $0.22381050036^0$ | 0.00390625 |
| $0.11190567706^0$ | 0.001953125 |
| $0.05595289189^0$ | 0.0009765625 |
| $0.02797645261^0$ | 0.00048828125 |
| $0.01398822714^0$ | 0.000244140625 |
| $0.00699411367535^0$ | 0.0001220703125 |
| $0.0034970568507^0$ | 0.00006103515625 |
| $0.00174852842698^0$ | 0.000030517578125 |
| $0.000874264213694^0$ | 0.0000152587890625 |
| $0.000437132106872^0$ | 0.00000762939453125 |
| $0.000218566053439^0$ | 0.00000381469726563 |
| $0.00010928302672^0$ | 0.00000190734863281 |
| $0.0000546415133601^0$ | 0.000000953674316406 |
| $0.00002732075668^0$ | 0.000000476837158203 |
| $0.00001366037834^0$ | 0.000000238418579102 |
| $0.00000683018917001^0$ | 0.000000119209289551 |
| $0.00000341509458501^0$ | 0.0000000596046447754 |
| $0.0000017075472925^0$ | 0.0000000298023223877 |
| $0.000000853773646252^0$ | 0.0000000149011611938 |
| $0.000000426886823126^0$ | 0.00000000745058059692 |
| $0.000000213443411563^0$ | 0.00000000372529029846 |
| $0.000000106721705781^0$ | 0.00000000186264514923 |
| $0.0000000533608528907^0$ | 0.000000000931322574615 |
| $0.0000000266804264454^0$ | 0.000000000465661287308 |

### 3.1. Basic CORDIC Processor

The internal configuration of a CORDIC processor is shown in Fig. 1. It consists of two functional modules: (i) the CORDIC operation module and (ii) the angle updating module. The CORDIC operation module performs CORDIC iterations specified in eq. (3). It consists of dual barrel shifters and dual adders to facilitate the updating of both $x_i$ and $y_i$ simultaneously. The number of bits to be shifted is controlled by the shift sequence $\{s(m,i)\}$, generated on chip with a simple counter and additional control devices. The add/subtract operation is determined by the sequence $\{\delta_i\}$. The angle updating module updates the rotational angles through simple addition operations specified in eq. (4). If $t_0$ be the time needed for performing a single CORDIC iteration then the total computation delay of a CORDIC block, $t_C$, is equal $M' t_0$. For circular rotation $M' = M$, but for hyperbolic rotation $M' > M$ (due to the requirement of some double pass operations). Here $t_0 = t_a + t_S$, $t_a$ and $t_S$ are the time needed to perform one addition and one shifter operations respectively. A lookup table is required to hold the elementary angles.

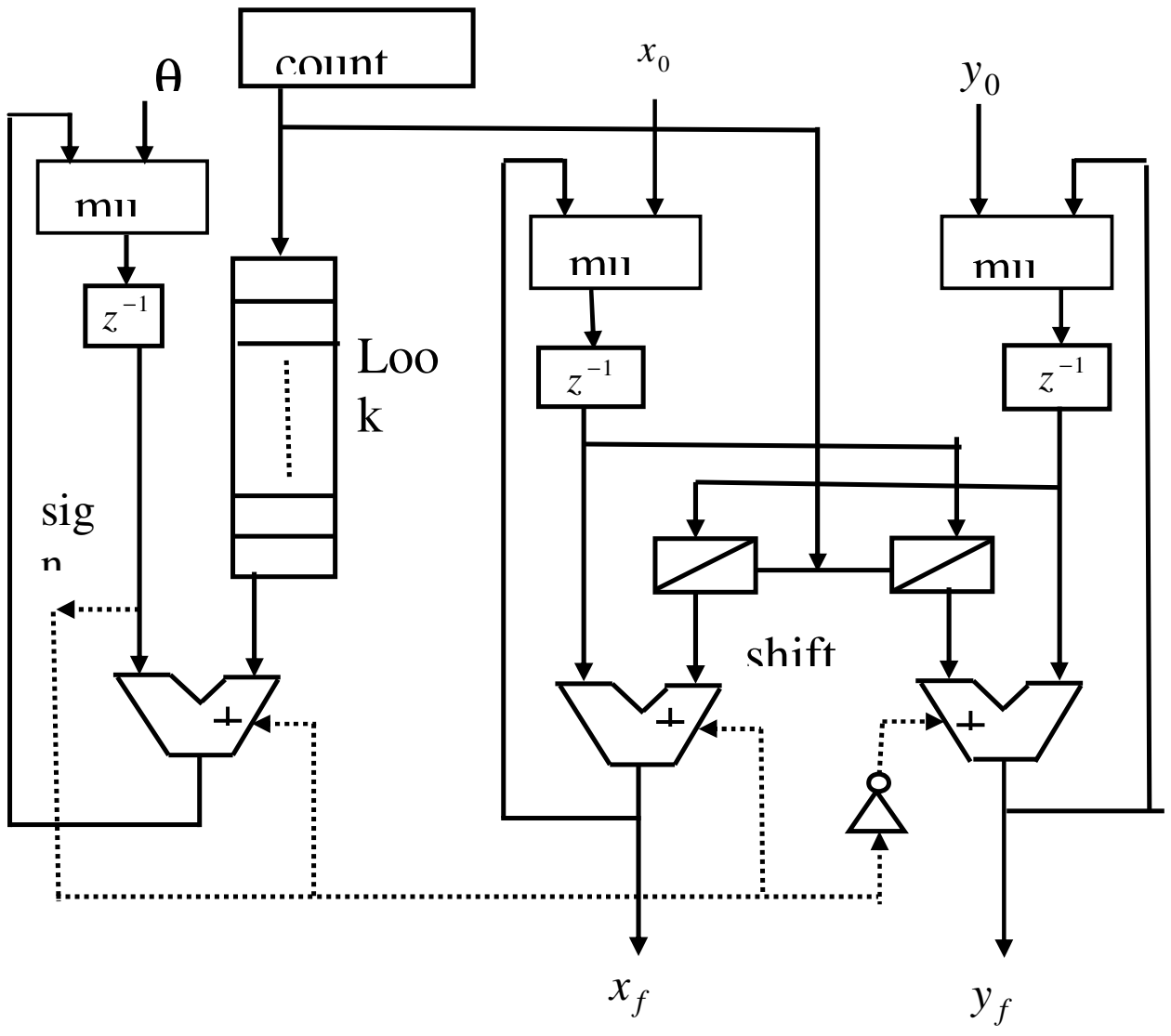

Figure 1. CORDIC block

## 4. THE TRIGONOMETRIC LMS ALGORITHM

In TLMS based adaptive filter, the *k*-th weight is expressed as $w_k = A_k \sin\theta_k$, $-\frac{\pi}{2} < \theta_k < +\frac{\pi}{2}$ for $k = 0, 1, \ldots, N-1$ (*N* is number of taps) and $A_k$ is a member of a set of selected *N* positive numbers so that the minima of the error performance surface is contained in the hypercube with vertices: $[\pm A_0, \pm A_1, \ldots, \pm A_{N-1}]^t$. Instead of updating weight, $\theta$ is updated in this algorithm. For an input $\mathbf{x}(n) = [x(n), x(n-1), \cdots, x(n-N+1)]^t$, the output and updatation equation of TLMS algorithm can be given by,

$$y(n) = \sum_{k=0}^{N-1} x(n-k) A_k \sin\theta_k(n), \qquad (21)$$

and $$\boldsymbol{\theta}(n+1) = \boldsymbol{\theta}(n) + \mu\, \Delta(n)\mathbf{x}(n)e(n), \qquad (22)$$

where, $\mu$ is the stepsize parameter, $\Delta$ is a diagonal matrix with $j$-th diagonal entry given by $\Delta_{j.j} = A_j \cos\theta_j$, $j$ = 0, 1,…,$N$-1, and error term $e(n) = z(n) - y(n)$, during initial training period $z(n) = d(n)$ (desired output) and subsequently it equates with the decision $\hat{y}(n)$ ($\hat{y}(n) = Q[y(n)]$).

For the hardware implementation, we may set $A_k = 1$ for all $k$, $k$=0,1,…,$N$-1, without loss of generality, since the hypercube $[\pm\frac{A_0}{A_r}, \pm\frac{A_1}{A_r}, \ldots, \pm\frac{A_{N-1}}{A_r}]^T$ normalized with $A_r = \max(|A_0|, |A_1|, \ldots, |A_{N-1}|)$ will always be contained within $[\pm1, \pm1, \cdots, \pm1]^T$. The normalization factor $A_r$ can be clubbed with the gain of the front-end amplifier. Therefore, in the present context, the output and update expressions of TLMS reduce to the following forms:

$$y(n) = \sum_{k=0}^{N-1} x(n-k) \sin\theta_k(n), \tag{23}$$

$$\theta_k(n+1) = \theta_k(n) + \mu e(n) \cos\theta_k(n) x(n-k). \tag{24}$$

For the HLMS algorithm, eq.(23) and (24) express in the following way:

$$y(n) = \sum_{k=0}^{N-1} x(n-k) \sinh\theta_k(n) \tag{25}$$

$$\theta_k(n+1) = \theta_k(n) + \mu e(n) \cosh\theta_k(n) x(n-k) \tag{26}$$

The architecture of the TLMS and the HLMS based adaptive filter is shown in Fig. 2.

## 5. SIMULATIONS

For the simulation of the TLMS and the HLMS algorithms software based model of the TLMS and the HLMS based adaptive filters have been developed using C programming language. Each of these consists of three functional modules:

i) **Module to generate transmitted signal**: Here, digital signals, which are discrete in time and discrete in amplitude, are generated. The transmitted symbols were chosen from an alphabet of 16 equispaced, equiprobable discrete amplitude levels with transmitted signal power of 10 dB.

ii) **Module for modeling the channel**: In this paper, we present simulation results for filtering an AWGN channel with transfer function $H(z) = (1+2z^{-1})(1-0.5z^{-1})(1+1.1z^{-1})(1-0.6z^{-1})$ and noise variance 0.077.

iii) **Filtering module**: This module is used to recover the transmitted signal. A 15 tap equalizer with center placed at the 8-th tap position was used for equalizing the channel and a step size of $\mu = .0004$ was adopted for theta adaptation. For each updated value of theta this module calls a submodule, which computes sine and cosine values using 16 or 32 steps of CORDIC rotations.

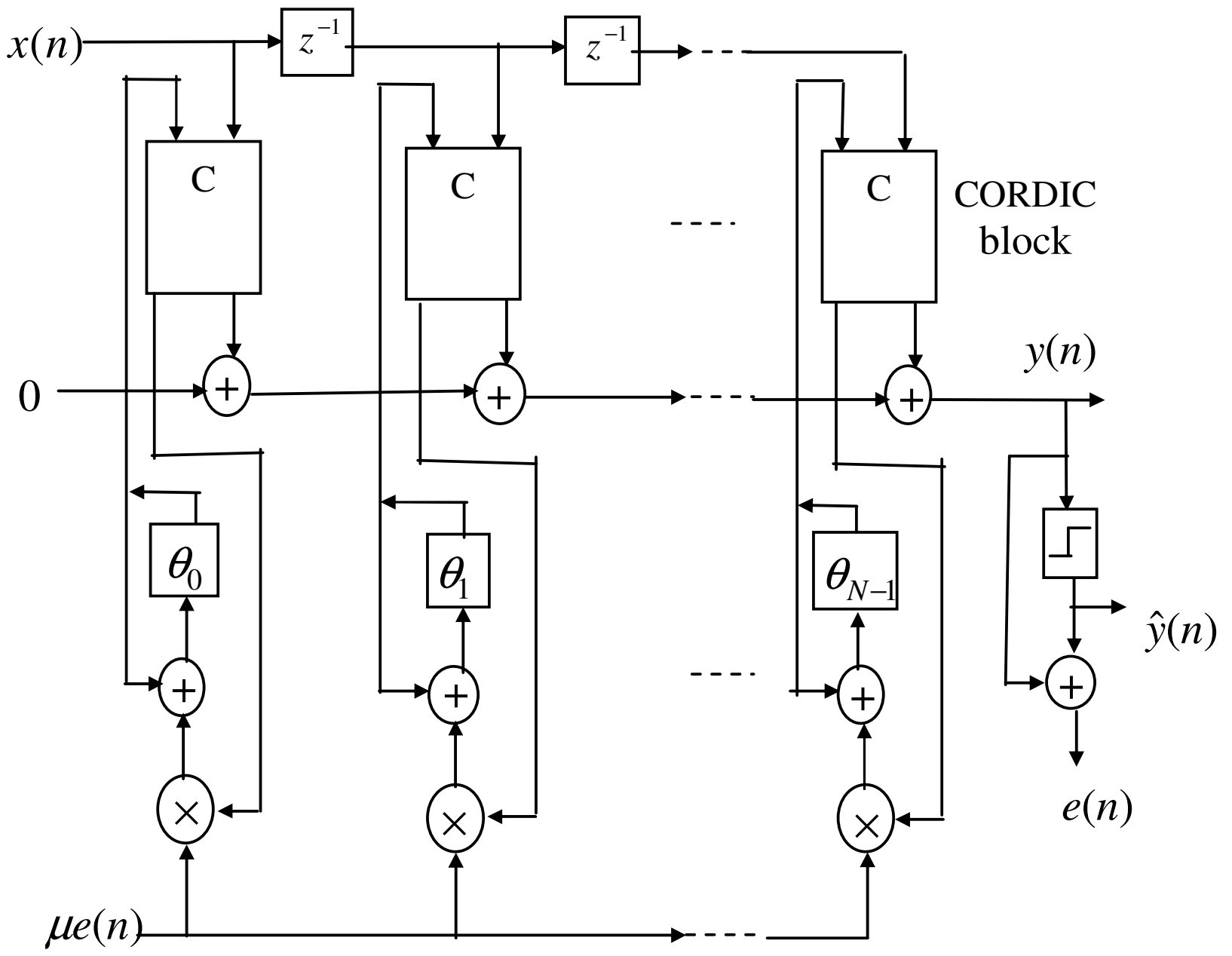

Figure 2. CORDIC based adaptive filter

Simulation result for the TLMS algorithm using 16 internal CORDIC rotations is presented in Fig. 3. The convergence performance of the same algorithm is observed using 32 steps CORDIC rotations that is given in Fig. 4. In Fig. 5, the convergence characteristic of the HLMS method is verified utilizing 32 steps rotations.

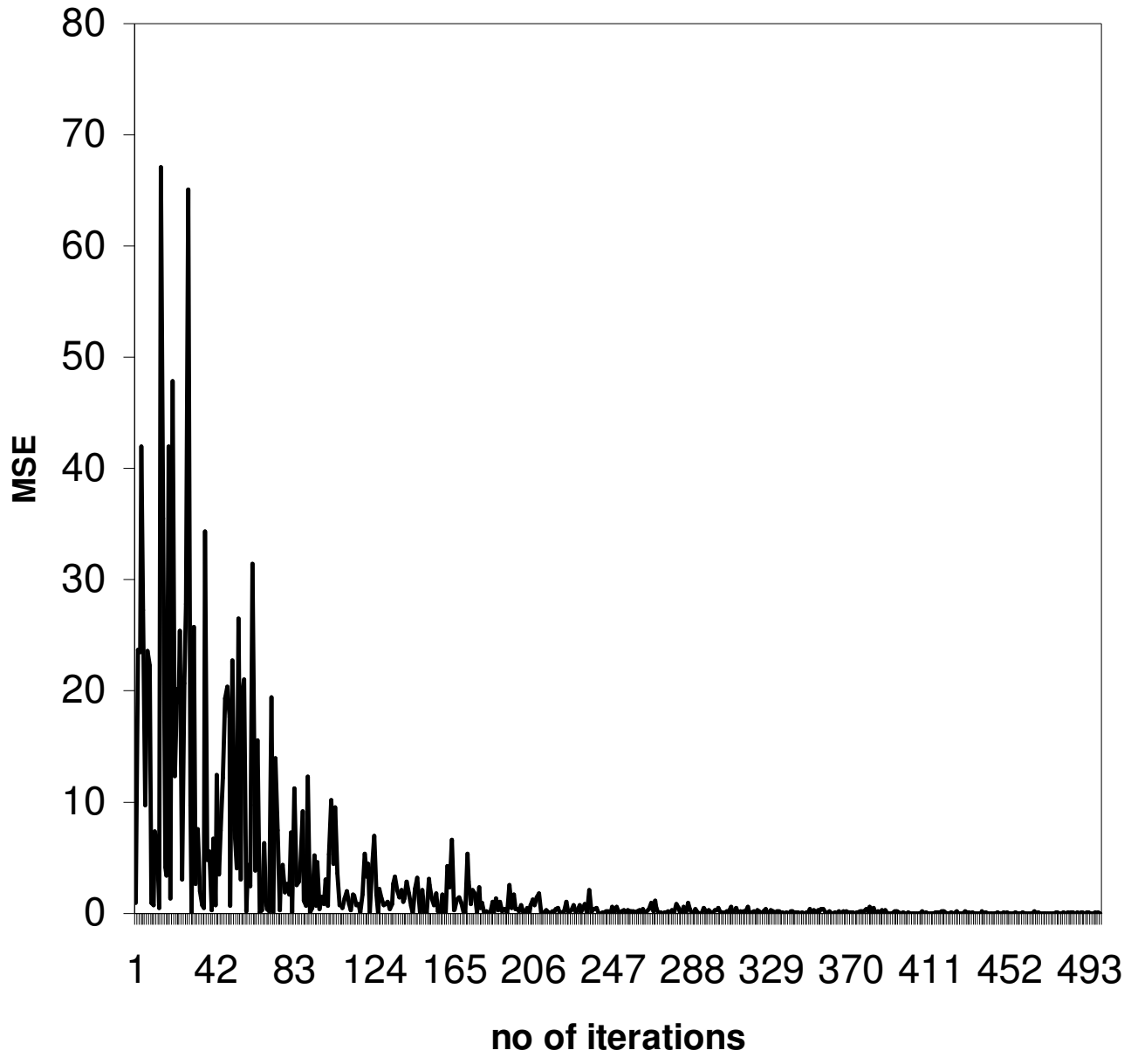

**Figure 2. TLMS using 16 rotations**

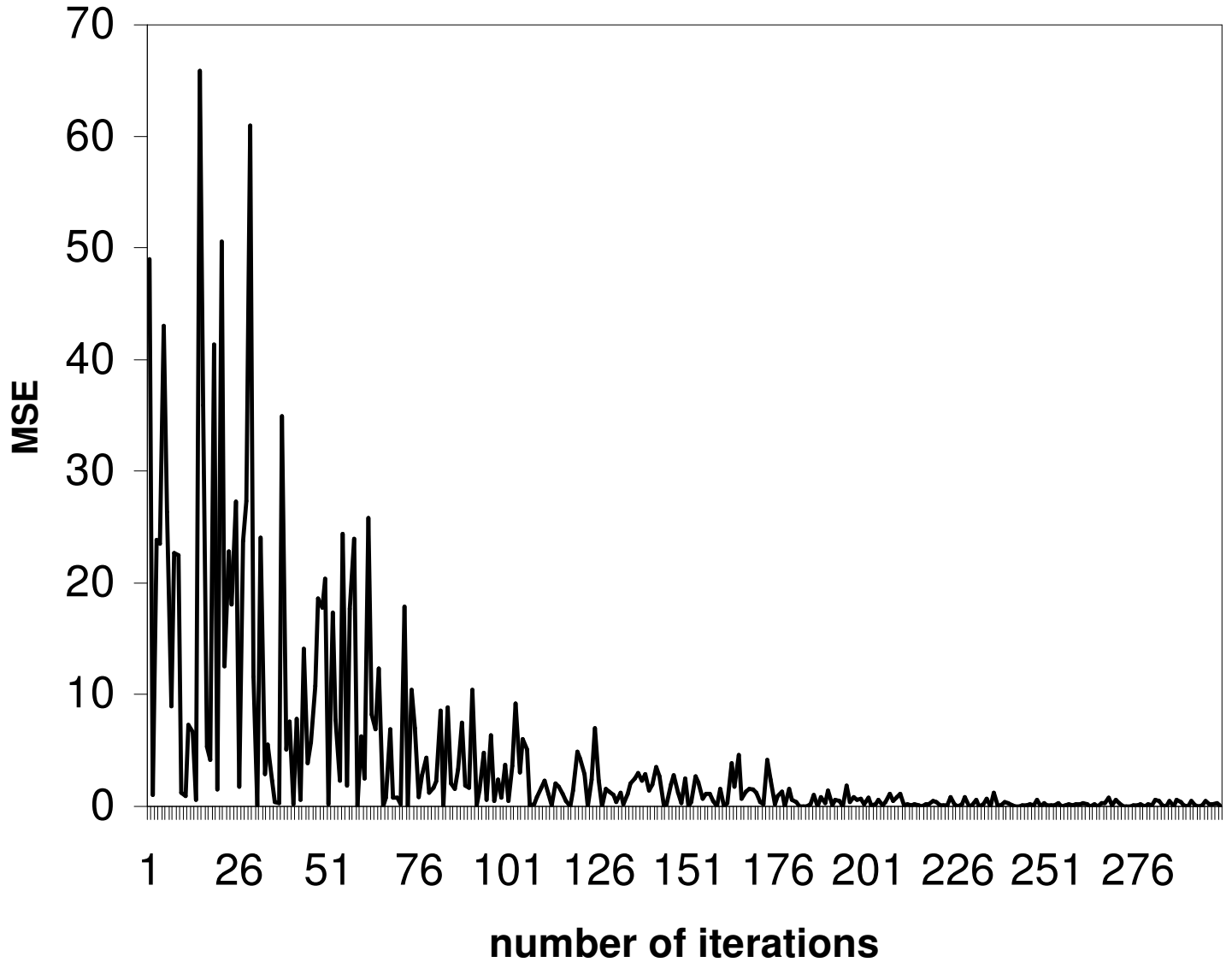

**Figure 4. TLMS using 32 rotations**

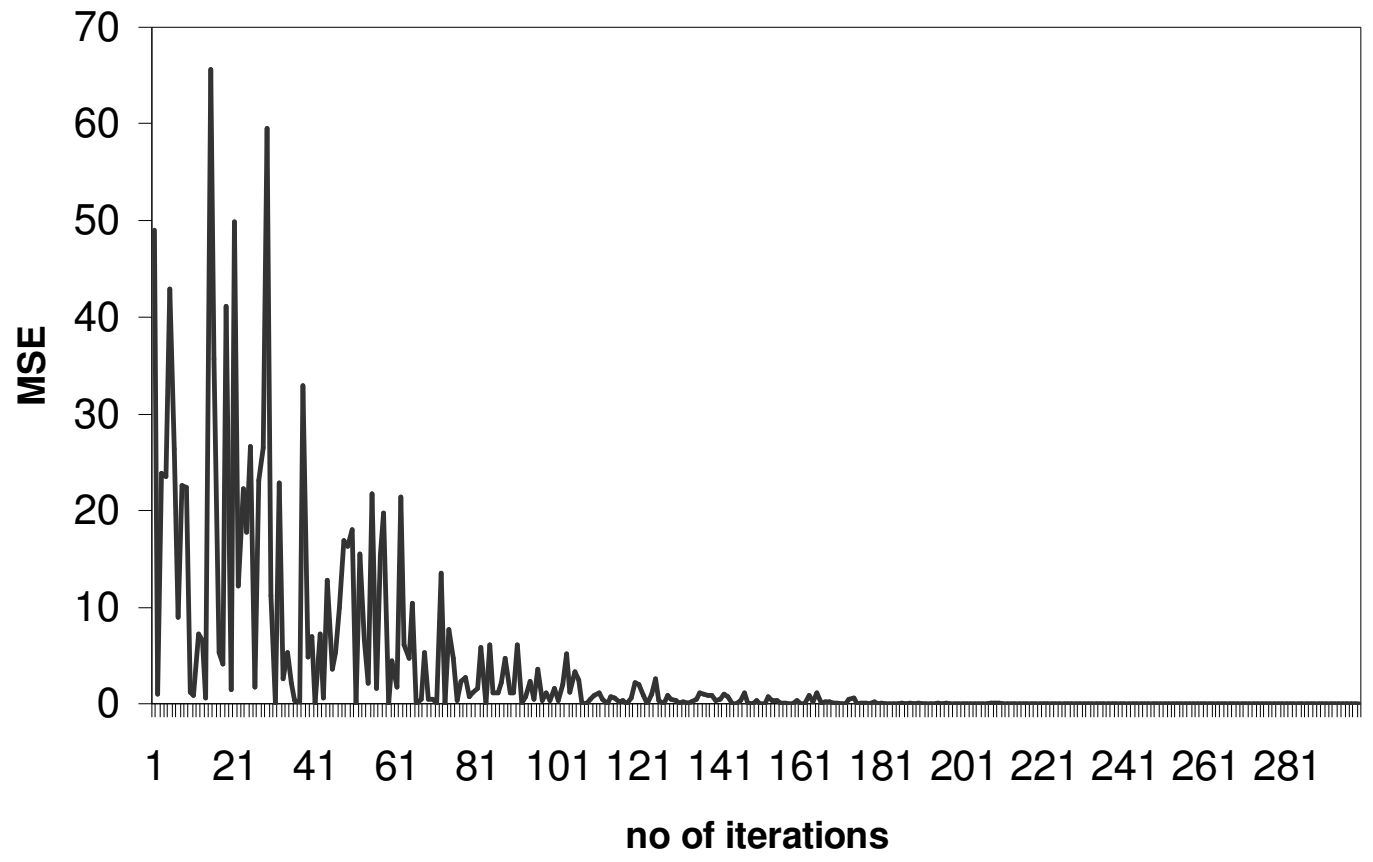

**Figure 5. HLMS using 32 rotations**

## 6. CONCLUSIONS

Adaptive filter is an important part of DSP. An adaptive filter is evaluated by certain criteria. Most important ones of those criteria are robustness, convergence, hardware complexity, cost of implementation, effectiveness etc. Several adaptive filtering algorithms have been proposed to meet the criteria. Among those, LMS [15] is the most widely used algorithm. Trigonometric LMS and Hyperbolic LMS are the two new versions of the LMS algorithm which can be efficiently implemented by CORDIC [3][5]. Although numerous research works have been done on the CORDIC algorithm and its use in implementation of the popular LMS algorithm, none of those works focused on the internal steps performed by the CORDIC processor. The convergence performance of the TLMS [1] algorithm and HLMS [2] algorithm was observed using CORDIC in some works. But the actual step-by-step rotations performed by the CORDIC hardware were never considered in any of those research works. Only the final and direct output of CORDIC block was used in implementation of the TLMS and HLMS algorithm. Though the results , that is, the convergence performance of the two algorithms using CORDIC block output directly were quite satisfactory , still there remains a question ,does the CORDIC block gives the same or nearly the same satisfactory result if its actual hardwired shifts and adds, namely the steps are considered in implementation? Our present work tried to solve the question. Previous works on implementation of the TLMS and HLMS algorithm using direct CORDIC block output were studied. Extensive works are done to simulate the internal step-by-step rotations of the CORDIC algorithm which will actually be followed in hardware. Finally the TLMS and HLMS algorithms are implemented using the actual steps, that is the shifts and adds performed by the CORDIC, and what we achieved was intensively satisfactory. The performances of these two algorithms were highly convincing. So we can conclude that the TLMS and HLMS implementation using CORDIC processor will be profoundly efficient.